\documentclass[fleqn,usenatbib]{mnras}

\usepackage{newtxtext,newtxmath}

\usepackage[T1]{fontenc}

\DeclareRobustCommand{\VAN}[3]{#2}
\let\VANthebibliography\thebibliography
\def\thebibliography{\DeclareRobustCommand{\VAN}[3]{##3}\VANthebibliography}

\usepackage{graphicx}	
\usepackage{amsmath}	

\title[Parameter free prediction]{Parameter-Free Prediction of the Asymptotic Acceleration Scale Confirmed by Weak Lensing}

\author[M.H.P.M. van Putten]{
Maurice H.P.M. van Putten$^{1,2}$\thanks{mvp@sejong.ac.kr}
\\
$^{1}$Physics and Astronomy, Sejong University, Seoul\\
$^{2}$INAF-Osservatorio Astronomico di Capodimonte, Salita Moiariello 16, I-80131 Napoli, Italy
}

\date{Accepted XXX. Received YYY; in original form ZZZ}

\pubyear{\the\year{}}

\begin{document}
\label{firstpage}
\pagerange{\pageref{firstpage}--\pageref{lastpage}}
\maketitle

\begin{abstract}
The de Sitter transition scale $a_{dS}=cH$ to anomalous galaxy dynamics was previously derived from first principles based on the background Hubble expansion $H$ and the velocity of light $c$. 
It introduces a $C^0$-transition across $a_{dS}$, 
which is in tension with predictions of $\Lambda$CDM 
galaxy models at a significance of $6\sigma$. 
Tracing late-time cosmology, it predicted the deep-asymptotic acceleration scale $a_0^{th}\simeq 1.66\times 10^{-8}{\rm cm\,s^{-2}}$ without adjustable parameters based on $H$ and the deceleration parameter $q$. 
A recent weak-lensing determination of $a_0^{WL} = 1.63_{-0.20}^{+0.23}\times 10^{-8}{\rm cm\,s}^{-2}$ now provides an independent empirical test of this prediction. 
The agreement with the updated theoretical value $a_0^{th} = 1.63_{-0.14}^{+0.13}\times 10^{-8}{\rm cm\,s^{-2}}$ closes the logical chain between cosmology, galaxy dynamics, and asymptotic scaling. 
Unlike phenomenological fits to rotation curves, this result does not rely on tuning or baryonic modeling but follows directly from background cosmology. This parameter-free model suggests some further tests by galaxy surveys extending over a finite range of redshifts.
\end{abstract}


\section{Introduction}

Galaxy dynamics exhibit a universal scaling that has long suggested the presence of a fundamental scale of acceleration, parameterizing the relation between photometric-spectroscopic  data on baryonic mass $M_b$ and orbital velocity $V_c$, described by the baryonic Tully-Fisher relation (bTFR).
It provides one of the cleanest empirical probes of galaxy dynamics in the weak-acceleration regime, where observed rotation curves deviate from Newtonian expectations based solely on baryonic mass $M_b$
\citep{tul77,mac12,fam12,lel16a,lel16b,div25}. 
At large orbital radii, $V_c$ tends to asymptote to a constant value, satisfying
\citep{mac12}
\begin{eqnarray}
M_b = A V_c^4,
\label{EQN_bTFR}
\end{eqnarray}
where $A=\left( 47\pm 6\right) M_\odot$ $\left({\rm km\,s^{-1}}\right)^{-4}$.  

Recently, (\ref{EQN_bTFR}) has been probed by direct measurements of total gravitational masses through weak gravitational lensing \citep{mis24}, confirming bTFR in the asymptotic regime of flat rotation curves. 
This effectively circumvents some of the uncertainties in previous photometric based rotation curve measurements, 
by direct measurement of total gravitating mass and sufficiently far out for stellar orbits to probe the gravitational potential of the host galaxy in effectively spherical symmetry.

The bTFR normalization is commonly expressed by an equivalent acceleration scale $a_0$ \citep{mil83a,mil83b,mil83c}, 
satisfying $GAa_0=1$, where $G$ is Newton's constant. The corresponding bTFR normalization satisfies \citep{mis24,mon25}
\begin{eqnarray}
    a_0^{WL}=1.63^{+0.23}_{-0.20}\times 10^{-8}{\rm cm\,s^{-2}}.
    \label{EQN_obs}
\end{eqnarray}
This weak-lensing result is significantly greater than conventional estimates from galaxy rotation curves \citep{mac12,fam12} 
\begin{eqnarray}
a_0^{RC}=\left(1.3\pm 0.3\right)\times 10^{-8}{\rm cm\,s^{-2}}
\label{EQN_12}
\end{eqnarray}
derived previously from rotation curves extracted from photometric-spectroscopic surveys of galaxy rotation curves,
corrected for asphericity of their Newtonian gravitational potentials. To some extent, this discrepancy may also be 
attributed to the need for an extrapolation, when rotation curves inadequately extend into the large distance 
asymptotics \citep{van18b}.
It has long since been recognized as comparable to $cH_0$ \citep{fam12}, i.e. the background de Sitter scale of acceleration \citep{van17b}
\begin{eqnarray}
    a_{dS}=cH,
    \label{EQN_adS}
\end{eqnarray}
where $H=H_0$ is the present Hubble parameter.

Remarkably, the value $a_0^{th}\simeq 1.66\times 10^{-8}{\rm cm\,s}^{-2}$ consistent with (\ref{EQN_obs}) 
was derived previously from background cosmology \citep{van17b}, evaluated for late-time cosmological parameters consistent with the Local Distance Ladder (LDL), prior to the present weak-lensing determinations. 

The new measurement provides an independent observational test of the cosmological origin of (\ref{EQN_bTFR}) through parameterization of anomalous galaxy dynamics by $a_{dS}$ and $a_0$ \citep{van18a}. This sensitivity to background cosmology goes beyond $\Lambda$CDM galaxy models. 
Indeed, the {\it Spitzer} Photometry Accurate Rotation Curves (SPARC) data indicate an apparent $C^0$-transition across (\ref{EQN_adS}), whereas the  
McMaster Unbiased Galaxy Simulations 2 (MUGS2) \citep{kel17} $\Lambda$CDM galaxy models exhibit inherently
smooth transitions. The resulting discrepancy is at the level of approximately $6\sigma$ \citep{van18a}. 
This cannot be removed by a simple rescaling of model parameters, and indicates a structural mismatch between 
the CDM paradigm of $\Lambda$CDM and the dynamics of the galaxy on these scales. 

Support for this framework requires a systematic assessment of 
\begin{eqnarray}
a_0=a_0(H,q)
\label{EQN_a0T}
\end{eqnarray}
based on precision measurements of (\ref{EQN_bTFR}) and precision cosmology, including the deceleration
parameter $q$.

The lensing-based estimate (\ref{EQN_obs}) adopts $H_0 = 73\,{\rm km\,s^{-1}\,Mpc^{-1}}$ and $\Omega_{M,0} = 0.2793$, 
consistent with analyzes of the  SPARC catalog \citep{lel16a}, measurements based on bTFR \citep{scho20,bro21} and WMAP \citep{hin13}. 
The chosen $H_0$ agrees with the Local Distance Ladder measurement $H_0 = 73.04 \pm 1.04\,{\rm km\,s^{-1}\,Mpc^{-1}}$ \citep{rie22}, 
while the corresponding density of physical matter $c_M = \Omega_{M,0} h^2$ for WMAP, $c_M^W = 0.1488$, 
is slightly higher $(\sim 10\%)$ than the Planck value $c_M^P = 0.1430 \pm 0.0011$ \citep{agh20}. 
This is expected to have negligible effect on $a_0$, since its dependence on $\Omega_{M,0}$ propagates only through $q_0$. In particular, the implied shift in $q_0$ $(\sim 10\%)$ is considerably smaller than its current observational uncertainty $(\sim 30\%)$ \citep{cam20}.

These parameter choices in (\ref{EQN_obs}) effectively anchor the analysis in late-time cosmology and away from a strict Planck–$\Lambda$CDM interpretation. Whereas standard $\Lambda$CDM predicts $q_0 \simeq -0.52$ \citep{agh20,har18}, late-time data indicate $q_0 \simeq -1$ \citep{van17b,cam20,abc21,des25}, supporting a dynamically accelerating background consistent with LDL measurements of $H_0$. 

In light of the recent weak-lensing determination (\ref{EQN_obs}), we revisit the potential cosmological origin of bTFR derived in \cite{van17b}. 
The new measurement, lying noticeably above conventional rotation-curve estimates, provides an opportunity to reassess the cosmological parameterization of the Milgrom scale (\ref{EQN_a0T}). 
This appears particularly timely in view of current tensions in cosmology \citep{div25}.

To study this, we consider the asymptotic scaling in (\ref{EQN_bTFR}) in light of the robust weak-lensing constraint $a_0^{WL}$ on the parameter-free prediction $a_0^{th}$, while preserving consistency with LDL and the Planck value of $c_M$. 

We first revisit a principal indeterminacy of inertia in the bTFR (\S2), 
allowing sensitivity of galaxy dynamics to background cosmology through reduced inertia without modification of general relativity. 
This interpretation preserves the geometric embedding of Newtonian gravity in classical general relativity, including standard Newtonian potentials and gravitational lensing by baryonic matter. 
With no free parameters, the resulting curvature-regulated inertia can be applied to late-time cosmology with dynamical dark energy. This 
yields a self-consistent parameterization compatible with LDL and the Planck constraint on the density of physical matter (\S3). 
We confront the resulting prediction for (\ref{EQN_a0T}) with the new lensing observation (\ref{EQN_obs}) in \S4 and summarize its implications in \S5.

\section{Indeterminacy of Inertia}

Galaxy dynamics may trace background cosmology rather than determine it. The cosmological background stands independently, 
constrained by early- and late-time observations, notably Planck CMB \citep{agh20} and large-scale structure surveys. 
Gravitational lensing is governed by null geodesics of general relativity, providing the covariant geometric embedding of the Newtonian potentials derived from the Poisson equation sourced by baryonic mass. This structure sets the gravitational structure in bTFR (\ref{EQN_bTFR}). 

Complementary to this structure is the inertial sector of stellar motion in (\ref{EQN_bTFR}).
The orbital motion of a star in a galaxy with total baryonic mass $M=M_b$ is parameterized by
\begin{itemize}
    \item Gravitating mass $m=m_{\tiny\rm gravity}$, parameterizing the Newtonian binding energy to the galaxy;
    \item Inertial mass $m^\prime=m_{\tiny\rm inertial}$, parameterizing orbital curvature according the Newton's second law of motion.
\end{itemize}
The Hamiltonian of circular orbital motion in the test particle limit $(m\ll M)$ is the total energy 
\begin{eqnarray}
{\cal H}(m,m^\prime) = E_k(m^\prime) + U_N(m)
\label{EQN_H}
\end{eqnarray}
defined by the sum of kinetic energy $E_k = \tfrac12 m^\prime \alpha r$ at centripetal acceleration $\alpha = V_c^2/r$ and Newtonian binding energy $U_N = -GMm/r$. 
Equating the Newtonian gravitational force with the observed orbital response implies the dynamical condition $m^\prime \alpha = m a_N$, where $a_N = GM/r^2$ is the expected Newtonian acceleration. Substitution into $E_k$ yields $E_k = {GMm}/{2r}$, independent of $m^\prime$. 
Since $U_N$ is fixed by the gravitating mass, the on-shell Hamiltonian is invariant with respect to inertia \citep{van17c,van24a}:
\begin{eqnarray}
    {\cal H}(m,m^\prime) \equiv {\cal H}_N(m),
    \label{EQN_HN}
\end{eqnarray}
where ${\cal H}_N(m)=-GMm/(2r)$ is the Newtonian Hamiltonian, familiar of planetary motion.

This ${\cal H}$-invariance with respect to inertia in (\ref{EQN_HN}) highlights that inertia is not determined by the gravitational sector. This indeterminacy extends to binaries with mass ratio $\eta=m_2/m_1$. 
In an equal-mass binary $(\eta = 1)$ or a test-particle limit $(\eta \to 0)$, the Hamiltonian remains invariant 
with respect to inertia, preserving the center-of-mass motion. 
For intermediate mass ratios $(0 < \eta < 1)$, the reduced mass $\mu = m_1 m_2/(m_1 + m_2)$ introduces a 
dependence of the relative kinetic energy on the partition of inertia; exact invariance then requires 
proportional changes in both inertial masses. 
In all cases, the gravitational potential is fixed which preserves the Newtonian binding energy, 
illustrating that inertia is not fixed by the gravitational sector.
An equivalent one-body representation is therefore straightforward for the limiting cases $\eta = 1$ and 
$\eta \to 0$, but less well-defined for intermediate $0 < \eta < 1$, where the partition of inertia affects 
the relative kinetic energy.
For the bTFR, the relevant limit is effectively $\eta \to 0$ of (\ref{EQN_H}-\ref{EQN_H0}),
corresponding to a test-particle orbiting a baryonic mass $M$. 

The on-shell invariance (\ref{EQN_HN}) indicates that inertia is not uniquely determined by the local gravitational sector. 
This structural indeterminacy allows a relational origin of inertia tied to the cosmological background, in the spirit of Mach’s principle \citep{mac93}. 
Mach’s original objection to Newton’s second law was precisely that inertia cannot be fixed locally by gravity alone; its origin must lie elsewhere, in relations with the mass distribution of the universe. 

The invariance exhibited here provides formal license to seek such closure in a nonlocal, cosmological origin of inertia.
The bTFR may represent precisely this closure in the low-acceleration regime below (\ref{EQN_adS}), encoding a cosmologically regulated reduction of effective inertia while gravity itself remains unmodified \citep{van17a}.

We next turn to a specific realization in conventional Friedmann cosmologies.

\section{A nonlocal origin of inertia}

In relativity, a particle rest mass $m=E/c^2$ is given by the Lorentz-invariant norm $E=\sqrt{-p^cp_c}$, locally defined by its four-momentum $p^b=mu^b$ along the tangent $u^b$ of its worldline.

\cite{mac93} proposed that inertial mass is relational and therefore inherently nonlocal.
For a particle at acceleration $a$, a Rindler horizon $h$ at a distance $\xi = c^2/a$ 
introduces a nonlocal extension of Einstein's equivalence principle \citep{rin60}. 
A nonlocal inertial mass is naturally identified
in the gravitational potential of $g=-a$ relative to $h$ \citep{van17b,van24c,van24a}, briefly revisited
in what follows.

In planetary systems, accelerations in orbital motions are sufficiently large for $h$ to be well within the cosmological horizon at $R_H$. 
Given an external force $F$, the {\it potential work integral}
\begin{eqnarray}
U_h = \int_0^h Fds
\label{EQN_U0}
\end{eqnarray}
then reproduces $mc^2$. 
This identification follows from the equivalence principle: in a uniformly accelerated frame, the Rindler horizon at $\xi=c^2/a$ defines the maximal causal domain. The work done against a constant force over this domain reproduces $mc^2$, providing a nonlocal representation of inertial mass equivalent to its local definition $E=mc^2$ (see \citealt{van17b,van24a} for detailed derivations).
By this equivalence, rest-mass energy admits both a local kinematic definition $m=E/c^2$ and a complementary nonlocal horizon-potential definition $m=U_h/c^2$ in its Rindler frame.

According to Newton's theory of gravitational attraction, particle trajectories about a gravitating mass $M$ derive from a Newtonian binding energy $U_N=-GM/r$ at separation $r$. Orbits about $M$ derive from a Newtonian gravitational force $F_N=ma_N$ in Newton's second law of motion, where $a_N= -U_N^\prime(r)$ is the expected Newtonian acceleration. Given an observed radial acceleration $a$, we infer an inertial mass $m^\prime$
by equating the Newtonian force with the observed orbital response,
\begin{eqnarray}
F_N\equiv ma_N = m^\prime a.
\label{eq:FN}
\end{eqnarray}
Here, $m$ denotes the mass entering the Newtonian gravitational force, while $m^\prime$ is the inertial mass inferred from the observed acceleration.
We interpret (\ref{eq:FN}) invariantly by the {\it expected} (Newtonian) and, respectively, {\it observed} orbital motion.

\subsection{Horizon-regulated transition}

In galaxy dynamics, however, $h$ readily drops beyond the cosmological horizon. In general, therefore, the causal floor in (\ref{EQN_U0}) is limited to the nearest horizon scale. Quite generally, therefore, 
\begin{eqnarray}
U\simeq \int_0^{\min\{\xi,R_H\}} F_Nds.
\label{EQN_Uh}
\end{eqnarray}
That is, $m^\prime = m$ for $\xi<R_H$, while  
$m^\prime = U/c^2 < U_h/c^2 = m$ by causal truncation
at $R_H$ whenever $\xi > R_H$. This introduces 
{\it reduced inertial} whenever $a<a_{dS}$,
leaving a non-Newtonian result
\begin{eqnarray}
U =  \int_0^{R_H} F_Nds \simeq  U_h \left(\frac{R_H}{\xi}\right) 
= U_h\left(\frac{a}{a_{dS}}\right) < mc^2.
\label{EQN_Uhp}
\end{eqnarray}
The relation $m^\prime/m = a/a_{dS}$ follows directly from the truncation of the integration domain at $R_H$, without introducing additional assumptions. 
What follows is Milgrom's law
$a \simeq \sqrt{a_{dS} a_N}$ in the approximation $a_0 = a_{dS}$, equivalently,
$\beta^2 = \sqrt{R_G/R_H}$ for $\beta=V_c/c$,
where $R_G=GM/c^2$ is the gravitational radius of a galaxy.

The transition between the Newtonian and cosmological regimes occurs at $\xi = R_H$, corresponding to $a=a_{dS}$. This transition does not intruduce a modification of Newtonian dynamics but rather expresses a consistency
condition imposed by horizon causality when $h$ drops beyond the cosmological horizon. According to this horizon crossing, the transition is expected to be relatively sharp. In fact, horizon crossing of $h$ at $R_H$ predicts a sharp $C^0$-transition to anomalous galaxy dynamics at accelerations below the de Sitter scale (\ref{EQN_adS}) when the two coalesce. 
Indeed, a gap at $\sim 6\sigma$ is apparent 
between SPARC and $\Lambda$CDM-galaxy models across $a_{dS}$ \citep{van18a}. This transition occurs at a radius $r_t\simeq \sqrt{R_HR_G}={\cal O}({\rm kpc})$.

\subsection{Curvature-regulated asymptotics}

bTFR (\ref{EQN_bTFR}) probes the asymptotic behavior for $a\ll a_{dS}$ beyond the cosmological horizon crossing scale in (\ref{EQN_Uhp}).
In this regime, orbiting particles probe the global structure of spacetime rather than the local Rindler or cosmological horizon. The relevant invariants are those governing null congruences in causal phase space, described by the Raychaudhuri equation.  

The asymptotics of $\xi\gg R_H$ should not introduce a new scale however. This naturally brings in the curvature scalar governing conformally invariant wave propagation. governed by the curvature scalar $J=(1-q)H^2/c^2$, the trace of the Schouten tensor
\citep{van24a}. This choice is {\it minimal}.
Once the (potential) work integral (\ref{EQN_Uh}) becomes sensitive only to background curvature, $\ell_J \sim J^{-1/2}$ now replaces the Rindler horizon rather than the cosmological horizon $R_H$ in the asymptotic regime.

To be specific, a three-flat FLRW cosmology consistent with LDL derives from a dynamical dark energy $\Lambda=J$
\citep{van15,van17b,van20}. 
With no free parameters, this predicts the scaling \citep{van25}
\begin{eqnarray}
H_0=\sqrt{\frac65}H_0^\Lambda=\left(73.79\pm 0.59\right){\rm km\,s^{-1}\,Mpc^{-1}}
\label{EQN_H0}
\end{eqnarray}
of the Planck-$\Lambda$CDM value $H_0^\Lambda = \left(67.36\pm 0.54\right)$ ${\rm km\,s^{-1}Mpc^{-1}}$ \citep{agh20}, 
consistent with the SH0ES value $H_0^{LDL}=\left(73.04\pm 1.04 \pm 1.04\right){\rm km\,s^{-1}Mpc^{-1}}$ \citep{rie22}.

Curvature-sensitivity of inertial mass on this background
is described by $a_0^{th}=(c^2/2\pi)\sqrt{J}$ \citep{van17b,van24a}, i.e.:
\begin{eqnarray}
a_0^{th} = \frac{\sqrt{1-q}}{2\pi}a_{dS}.
\label{EQN_a0J}
\end{eqnarray} 
In this framework, galaxy dynamics trace background cosmology parameterized by {\it two} distinct scales of acceleration:
the de Sitter scale (\ref{EQN_adS}), marking a horizon-scale transition to anomalous dynamics, and 
the relatively smaller asymptotic scale (\ref{EQN_a0J}) in (\ref{EQN_bTFR}) tracing background curvature rather than the geometric scale $R_H$. This fixes its parameterization in $a_0$ by $(H_0,q_0)$ of the cosmological background.

\section{Confrontation with Observations}

Evaluation of the prediction (\ref{EQN_a0J}) for late-time cosmological parameters of the previous section enables a direct confrontation with the observation (\ref{EQN_obs}).
Including $q_0\simeq -1.08\pm0.29$ of LDL \citep{cam20}, 
the model predicts 
\begin{eqnarray}
    a_0^{th}= 1.63^{+0.13}_{-0.14}\times 10^{-8}{\rm cm\,s^{-2}}
    \label{EQN_th}
\end{eqnarray}
in agreement with (\ref{EQN_obs}).
The uncertainty is deliberately conservative, obtained by propagating the up- and down errors in $H_0$ and $q_0$ without assuming any covariance or correlation between them. 
This ensures that the theoretical prediction is bounded by the widest plausible range given current cosmological measurements, while still yielding a central value in close agreement with the normalization (\ref{EQN_obs}).

The agreement $a_0^{th} \simeq a_0^{WL}$ lies within the observational uncertainties of both LDL and the bTFR normalization.
Derived from first principles without fine-tuning, this concordance suggests that the bTFR may originate in a finite sensitivity to background cosmology through the indeterminate origin of inertia. In this interpretation, gravitational lensing remains governed by general relativity. Sourced by baryonic matter alone, it does not invoke dark matter or modified gravity.
Specifically, bTFR normalization is not an independent empirical constant but is directly linked to background cosmology and, as such, is expected to evolve with cosmological redshift.

\begin{figure}
   \centering
   \includegraphics[width=\columnwidth]{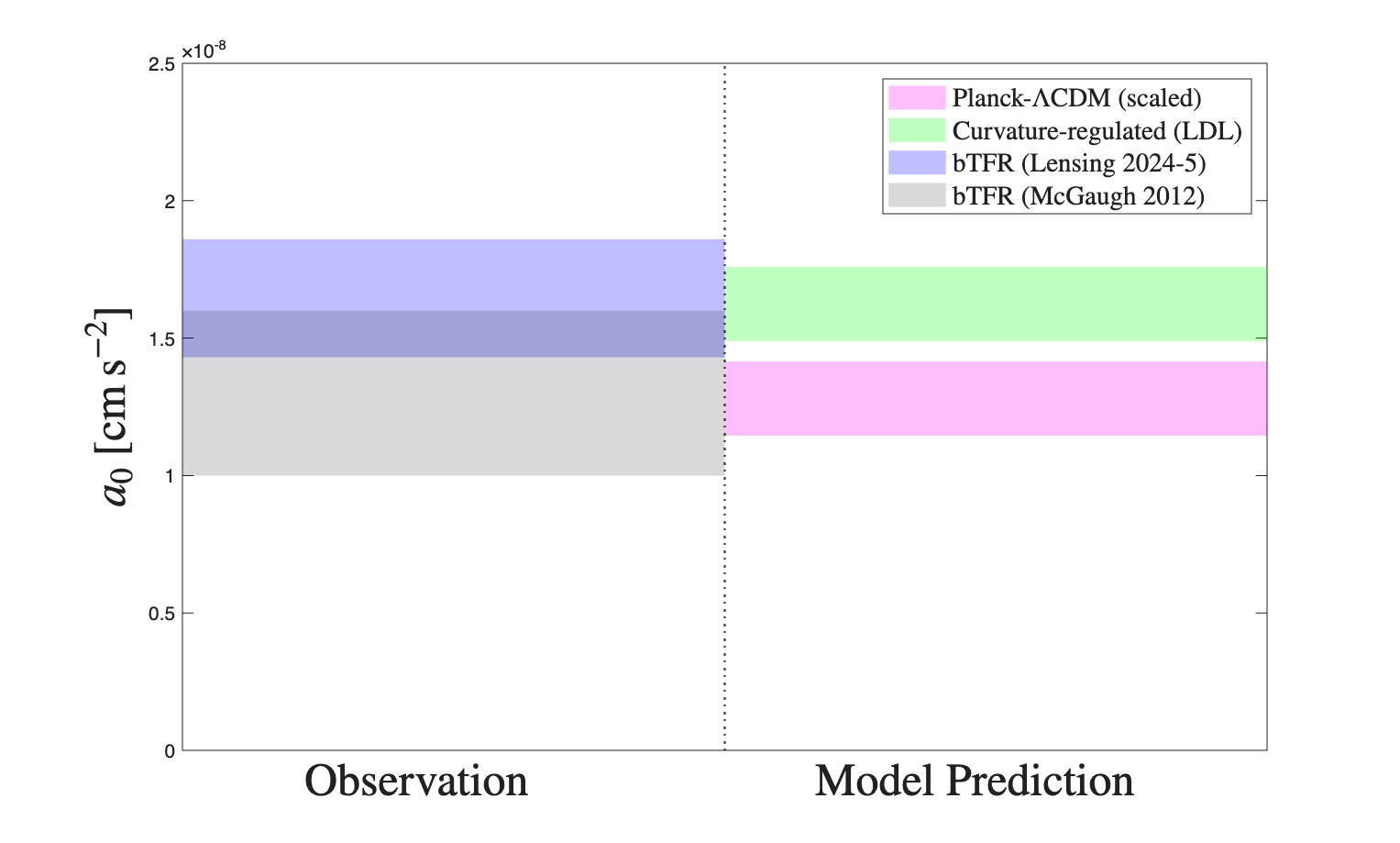}
   \caption{Observed and theoretical values of the acceleration scale $a_0$ in bTFR. Observational bands correspond to rotation-curve measurements of gas-rich galaxies \citep{mac12} and to recent weak-lensing determinations \citep{mis24,mon25}. Theoretical predictions are shown for curvature-regulated inertia in late-time cosmology with LDL parameters and for the Planck-$\Lambda$CDM concordance model. Band widths indicate $\pm1\sigma$ uncertainties. The theoretical prediction agrees with the weak-lensing normalization within observational uncertainties, while the Planck-based prediction lies systematically lower.}
   \label{F1}
\end{figure}

\section{Conclusions and outlook}

As a closure to the indeterminacy of inertia in orbital dynamics (\S2), 
Newton’s second law may be viewed as an energy-based relation, most clearly seen in a Rindler frame comoving 
along an orbit, 
\begin{eqnarray}
F_N \equiv - U_N^\prime = - U_h^\prime \equiv - \xi^{-1}U_h.
\label{eq:FU}
\end{eqnarray}
It implicitly introduces causality in the inertial sector, and is extended in \S3 to a cosmological background.
This balance appears naturally in the corresponding reductions of inertia ($m\rightarrow m^\prime$) and mass-energy at infinity. The latter is given by $E = mc^2\rightarrow E^\prime$ through the binding-energy deficit $U_N  = -E\,(R_G/r)$ -- encoded in spacetime curvature in general relativity without closure of the inertial sector.
It reveals the sensitivity of galaxy dynamics to the cosmological background. In weak gravitation, the two scales $R_G$ and $R_H$ appear side by side.
In the asymptotic regime, this sensitivity is governed by a curvature scale in the present framework and by the Hubble expansion rate in hyperconical gravity. This connection is consistent with the broader idea (often associated with Mach) that inertia reflects the global environment, here manifest in galaxy dynamics through the baryonic Tully–Fisher relation (\ref{EQN_bTFR}).

The agreement $a_{0}^{WL}\simeq a_0^{th}$ of weak-lensing observation and theory (\S4), reflecting the asymptotic sensitivity to background curvature, closes the logical chain between cosmology, galaxy dynamics, and asymptotic scaling. 
Unlike phenomenological fits to rotation curves, 
this result follows directly from background cosmology. 
It constitutes a robust parameter-free validation of the presented theory and specifically so for the transition scale (\ref{EQN_adS}) to anomalous galaxy dynamics.

The present conclusion reinforces earlier indications that there is no empirical need for dark-matter clustering on the scale of galatic disks \citep{van16a}, even if a cosmological dark-matter component is present. Agreement of the weak-lensing normalization with the baryonic Poisson equation sharpens this scale separation: any dark component must cluster on cosmological scales while remaining dynamically subdominant within the optical regions of galaxies.

For the Planck-$\Lambda$CDM concordance values $H_0^\Lambda$ and $q_0\simeq -0.52$, in contrast, the predicted normalization is significantly lower by a factor $\sqrt{8/5}\simeq 1.27$, as illustrated in Fig.~1.
The current lensing data therefore favor late-time cosmological parameters over the Planck concordance values in the context of curvature-regulated inertia. 

In light of the present agreement with weak-lensing data at $z=0$, this predicted evolution becomes a well-defined target 
for future measurements at intermediate and high redshifts, e.g., using MaNGA \citep{lee25}, to provide an independent observational tests of the theory. Alternatively, it may serve as a novel probe of dynamical dark energy. 

Specifically, we note that scaling by $a_{dS}$ alone would imply $\left.da_0/dz\right|_{z=0}>0$. The additional curvature dependence in (\ref{EQN_a0J}) through $q(z)$ reverses the sign, yielding $\left.da_0/dz\right|_{z=0}<0$
with the derivative approaching zero near $z\simeq 0.5$, reflecting the transition in cosmic acceleration. Extended to cosmic dawn, this sensitivity predicts anomalously early galaxy formation \citep{van24_pdu}.

\mbox{}\\
\centerline{\bf Acknowledgments}
The author thanks the anonymous reviewer for constructive comments. This research is supported, in part, by NRF RS-2024-00334550.

\section*{Data Availability}
This article uses only public data published in the
refereed literature.

\bibliographystyle{mnras}
\bibliography{bibfile} 

\bsp	
\label{lastpage}
\end{document}